\documentclass[a4paper]{article}

\usepackage{INTERSPEECH2019}
\usepackage{amsmath,graphicx}
\usepackage{booktabs}
\usepackage{microtype}
\usepackage{paralist}
\usepackage{hyperref}
\usepackage{cite}
\usepackage{amssymb}

\hypersetup{
    colorlinks=false,
    pdfborder={0 0 0},
}

\title{Parrotron: An End-to-End Speech-to-Speech Conversion Model and its Applications to Hearing-Impaired Speech and Speech Separation}

\name{Fadi Biadsy, Ron J. Weiss, Pedro J. Moreno, Dimitri Kanevsky, Ye Jia}
\address{Google}
\email{\{biadsy,ronw,pedro,dkanevsky,jiaye\}@google.com}

\begin{document}

\maketitle
\begin{abstract}
We describe Parrotron, an end-to-end-trained
speech-to-speech conversion  model that maps an input
spectrogram directly to another spectrogram, without utilizing any intermediate discrete
representation. The network is composed of an encoder, spectrogram and phoneme decoders, followed by a vocoder to synthesize a time-domain waveform.
We demonstrate that this model can be trained to normalize speech from any speaker regardless of
accent, %
prosody, %
and background noise, into the voice of a {\em single} canonical
target speaker with a fixed accent and consistent articulation and prosody.
We further show that this normalization model can be adapted to normalize highly atypical speech from a deaf speaker, resulting in
significant improvements in intelligibility and naturalness, measured via a speech recognizer and listening tests.
Finally, demonstrating the utility of this model on other speech tasks, we show that the same model architecture can be trained to perform a speech separation task.

\end{abstract}

\section{Introduction}

Encoder-decoder models with attention have recently shown
considerable success in modeling a variety of complex sequence-to-sequence problems.
These models have been successfully adopted to tackle a diverse set of tasks in
speech and natural language processing, such as machine translation
\cite{bahdanau2014neural}, speech recognition \cite{chan2016listen},
and even combined speech translation \cite{berard2016listen}.
They have also achieved state-of-the-art results
in end-to-end Text-To-Speech (TTS) synthesis~\cite{tac2} and Automatic Speech
Recognition (ASR)~\cite{chiu2018sota}, using a single neural network that directly
generates the target sequences, given virtually raw inputs.

In this paper, we combine attention-based speech recognition and synthesis
models to build a direct end-to-end speech-to-speech sequence transducer.
This model generates a speech spectrogram as a function of a different input spectrogram, with no
intermediate discrete representation.

We test whether such a unified %
model is powerful enough to normalize arbitrary speech from multiple accents, imperfections,  potentially
including background noise, and generate the same content in the voice of a {\em single} predefined target speaker.
The task is to project away all non-linguistic
information, including speaker characteristics, and to retain only what is being
said, {\em not} who, where or how it is said. This amounts to a text-independent, many-to-one
voice conversion task \cite{toda2007one}.
We evaluate the model on this voice normalization task using ASR and listening studies, verifying that it is able to preserve the underlying speech content and project away other information, as intended.

We demonstrated that the pretrained normalization  model can be adapted to perform a more challenging task of converting
highly atypical speech from a deaf speaker into fluent speech,
significantly improving intelligibility and naturalness.
Finally, we evaluate whether the same network is capable of performing a speech separation task.
Readers are encouraged to listen to sound examples
on the companion website.~\footnote{\url{https://google.github.io/tacotron/publications/parrotron}}

A variety of techniques have been proposed for voice conversion, %
including
mapping code books~\cite{Abe1988}, neural networks~\cite{Watanabe, Narendranath}, dynamic
frequency warping~\cite{Valbret}, and Gaussian mixture models~\cite{Kain, Toth, Toda}.
Recent work has also addressed accent conversion \cite{Bearman, Felps} .
In this paper we propose an end-to-end architecture that directly generates the target signal,
synthesizing it from scratch. %
It is most similar to recent work on sequence-to-sequence voice conversion \cite{haque2018conv,zhang2019sequence,tanaka2018atts2s}.
\cite{haque2018conv} uses a similar end-to-end model, conditioned on speaker identities, to transform
word segments from multiple speakers into multiple target voices.
Unlike \cite{zhang2019sequence}, which trained separate models for each source-target speaker pair, we focus on many-to-one conversion.
Our model is trained on source-target spectrogram pairs, without augmenting inputs with bottleneck features from a pretrained speech recognizer to more explicitly capture phonemic information in the source speech \cite{zhang2019sequence}.
However, we do find it helpful to multitask train the model to predict source speech phonemes.
Finally, in contrast to \cite{tanaka2018atts2s}, we %
train the model without auxiliary alignment or auto-encoding losses.

Similar voice conversion techniques have also been applied to improving intelligibility for speakers with vocal disabilities \cite{yamagishi2012speech, kain2007improving}, and hearing-impaired speakers in particular \cite{lee2006spectral}.
We apply more modern machine learning techniques to this problem, and demonstrate that, given sufficient training data, an end-to-end trained one-to-one conversion model can dramatically improve intelligibility and naturalness of a deaf speaker.

\section{Model Architecture}
\label{sec:arch}

We use an end-to-end sequence-to-sequence model architecture that
takes an input source speech and generates/synthesizes target speech as output.
The only training requirement of such a model is a parallel corpus of paired input-output speech utterances.
We refer to this speech-to-speech model as {\em Parrotron}.

\begin{figure}[t]
  \centering
  \includegraphics[width=\columnwidth]{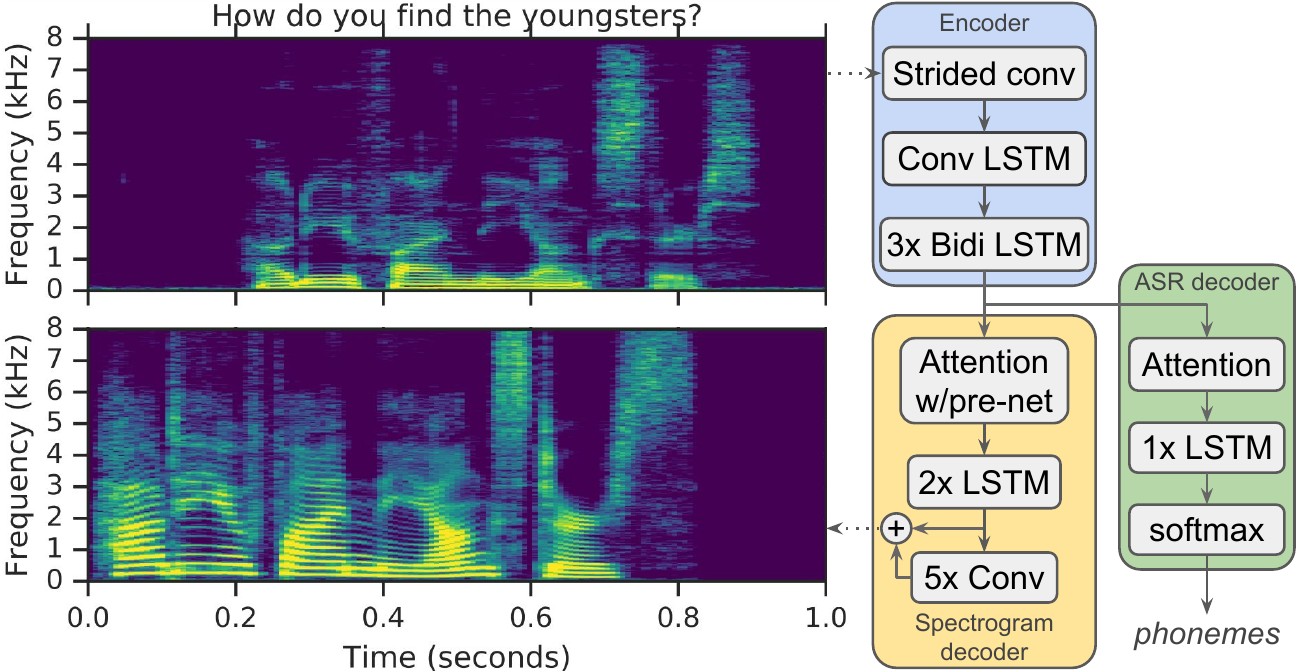}
  \caption{Overview of the Parrotron network architecture.
    The output speech is from a different gender (having higher pitch and formants), and has a slightly slower speaking rate.}
  \label{fig:normalization}
  \vspace{-0.2cm}
\end{figure}

As shown in Figure~\ref{fig:normalization}, the
network is composed of an encoder and a
decoder with attention, followed by a vocoder to synthesize a time-domain waveform.
The encoder converts a sequence of acoustic frames into
a hidden feature representation which the decoder consumes
to predict a spectrogram. The core architecture 
is based on recent attention-based end-to-end ASR models~\cite{chan2016listen,zhang2016very}
and TTS models such as Tacotron~\cite{tac2, tac1}.

\subsection{Spectrogram encoder}

The base encoder configuration is similar to the encoder in~\cite{weiss2017sequence},
and some variations are evaluated in Section~\ref{sec:norm}.
From the input speech signal, sampled at 16~kHz, we extract 80-dimensional log-mel spectrogram features
over a range of 125-7600 Hz, calculated
using a Hann window, 50~ms frame length, 12.5~ms frame shift, and 1024-point
Short-Time Fourier Transform (STFT).

The input features are passed into a stack of two convolutional layers
with ReLU activations, each consisting of 32 kernels, shaped
$3 \times 3$ %
in time $\times$ frequency, and strided by
$2\times2$, downsampling the sequence
in time by a total factor of 4,
decreasing the computation in the following layers.
Batch normalization~\cite{ioffe2015batch} is applied after each layer.

This downsampled sequence  is passed into a
bidirectional convolutional LSTM (CLSTM)~\cite{xingjian2015convolutional,schuster1997blstm} layer using
a $1\times3$ filter, i.e.\ convolving only across the frequency axis
within each time step. Finally, this is passed into a stack of
three bidirectional LSTM layers of size 256 in each direction,
interleaved with a 512-dim linear projection, followed
by batchnorm and ReLU activation, to compute the
final 512-dim %
encoder representation.

\subsection{Spectrogram decoder}

The decoder targets are
1025-dim STFT magnitudes, computed with the same framing as the input features,
and a 2048-point
FFT. %
We use the decoder network described in~\cite{tac2}, consisting of an
autoregressive RNN to predict the output spectrogram from the encoded input
sequence one frame at a time.
The prediction from the previous decoder time step
is first passed through a small pre-net containing 2 fully connected layers of
256 ReLU units, which was found to help to learn
attention \cite{tac1, tac2}. The pre-net output and attention context vector
are concatenated and passed through a stack of 2 unidirectional LSTM layers with 1024 units.
The concatenation of the LSTM output and the attention context vector is then
projected through a linear transform to produce a prediction of the target
spectrogram frame. Finally, these predictions  are passed through  5-layer
convolutional post-net which predicts a residual to add to the initial
prediction.  Each post-net layer
has 512 filters shaped $5\times1$ followed by batch normalization and tanh
activation.

To synthesize an audio signal from the predicted spectrogram,
we primarily use the Griffin-Lim algorithm~\cite{griffin1984signal} to estimate a phase %
consistent with the predicted magnitude, followed by an inverse STFT.
However, when conducting human listening tests we instead use a WaveRNN~\cite{kalchbrenner2018efficient} neural vocoder
which has been shown to significantly improve synthesis fidelity~\cite{tac2,jia2018transfer}.

\subsection{Multitask training with an ASR decoder}
\label{sec:multitask}

Since the goal of this work is to generate only speech sounds and not arbitrary
audio, jointly training the encoder network to simultaneously learn a high level
representation of the underlying language serves to bias the spectrogram decoder
predictions toward a representation of the same underlying speech content.
We accomplish this by adding an %
auxiliary ASR decoder to predict the
(grapheme or phoneme) transcript of the output speech, conditioned on the
encoder latent representation.
Such a multitask trained encoder can be thought of as learning a latent representation of
the input that maintains information about the underlying transcript, i.e.\ one
that is closer to the latent representation learned within a TTS sequence-to-sequence network.

The decoder input is created by
concatenating a 64-dim embedding for %
the grapheme emitted at
the previous step, and the 512-dim attention context. %
This is passed into a 256 unit LSTM layer. Finally the concatenation of the attention context
and LSTM output is passed into a softmax to predict
the probability of emitting each grapheme in the output vocabulary.

\section{Applications}
\label{sec:apps}

\subsection{Voice normalization}
\label{sec:norm}

We address the task of normalizing speech from an arbitrary speaker
to the voice of a predefined canonical speaker.
As discussed in Section~\ref{sec:arch}, to make use of Parrotron, we require a
parallel corpus of utterances spanning a variety of speakers and recording
conditions, each mapped to speech from a canonical speaker.
Since it is impractical to have single speaker record many hours of speech %
in clean acoustic environment, we use Google's Parallel WaveNet-based TTS~\cite{prodWavenet}
system to generate training targets from a large hand-transcribed speech corpus.
Essentially this reduces the task to reproducing any input speech in the voice of a single-speaker TTS system.
Using TTS to generate this parallel corpus ensures that:
\begin{inparaenum}[(1)]
\item the target is always spoken with a consistent predefined speaker and accent; %
\item without any background noise %
  or disfluencies.
\item Finally, we can synthesize as much data as necessary to scale to very large corpora.
\end{inparaenum}

\subsubsection{Experiments}

We train the model on a $\sim$30,000 hour training set consisting of about 24 million English
utterances which are anonymized and manually transcribed,
and are representative of Google's US English voice search traffic.
Using this corpus, we run a TTS system to generate target utterances in a synthetic female voice.

\begin{table}[t]
  \caption{
    \label{tbl:exps}
  WER comparison of different architecture variations combined with
  different auxiliary ASR losses.} %
  \vskip-0.2cm

  \centering
  \setlength{\tabcolsep}{1.0ex}
  \begin{tabular}{lcccc}
    \toprule
    ASR decoder target %
    & \hskip-1.5ex\#CLSTM\hskip-0.5ex & \hskip-0.5ex\#LSTM\hskip-1ex & Attention & WER \\
    \midrule
    None & 1 &  3 & Additive & 27.1  \\
    \addlinespace %
    Grapheme & 1 &  3 & Additive & 19.9 \\
    Grapheme & 1 &  3 & Location & 19.2 \\
    \addlinespace %
    Phoneme & 1 &  3 & Location & 18.5 \\
    Phoneme & 0 &  3 & Location & 20.9 \\
    Phoneme & 0 &  5 & Location & 18.3 \\
    \addlinespace %
    Phoneme w/slow decay \hspace{-1em} & 0 &  5 & Location & 17.6 \\
    \bottomrule
  \end{tabular}
  \vskip-0.3cm
\end{table}

To evaluate whether Parrotron preserves the
linguistic content of the original input signal after normalization,
we report word error rates (WERs) using a state-of-the-art ASR engine
on the Parrotron output as a measure of speech intelligibility.
Note that the ASR engine is not trained on Griffin-Lim synthesized speech,
a domain mismatch leading to higher WER. Table~\ref{tbl:exps} compares different architecture and loss configurations, evaluated on a hand-transcribed held-out test set of 10K anonymized utterances sampled from the same distribution as the train set.

The WER on the original speech (matched condition) is 8.3\%, which can be viewed as an upper bound.
Synthesizing the {\em reference} transcripts with a high quality TTS model and transcribing them using our ASR engine obtains a WER of 7.4\%.

The top row of Table~\ref{tbl:exps} shows performance using the base model architecture described in
Section~\ref{sec:arch}, using a spectrogram decoder %
employing additive attention~\cite{bahdanau2014neural} without an auxiliary ASR loss.
Adding a parallel decoder to predict graphemes %
leads to a significant improvement, reducing the WER from 27.1\% to
19.9\%.
Extending the additive attention with a location sensitive term
\cite{chorowski2015attention}
further improves results. %
This improves outputs on
long utterances where additive attention sometimes %
failed.

Since orthography in English does not uniquely predict %
pronunciation, we hypothesize that using phoneme targets for the ASR decoder
(obtained from forced alignment to the reference transcript)
may reduce noise propagated back to the encoder. Indeed
we find that this %
also shows consistent improvements. %

Turning our attention to the encoder architecture, we found that reducing
the number of parameters by removing the CLSTM significantly hurts
performance. However, using 2 extra BLSTM layers instead of the
CLSTM slightly improves results, while simultaneously simplifying the model.
Hyperparameter tuning revealed that simply using slower learning rate decay (ending at 90k instead of 60K steps) on our best model
yields 17.6\% WER. See Figure~\ref{fig:normalization} for an example model output.

\begin{table}[t]
\caption{
    \label{tbl:listening_test}
    Performance of Parrotron models on real speech.}
    \vskip-0.2cm
  \centering
  \begin{tabular}{lcr}
    \toprule
    Model & MOS & WER \\
    \midrule
    Real speech & 4.04 $\pm$ 0.19 & 34.2 \\
    \midrule

    Parrotron (female) & 3.81 $\pm$ 0.16 & 39.8 \\
    Parrotron (male) & 3.77 $\pm$ 0.16 & 37.5 \\
    \bottomrule
  \end{tabular}
\end{table}

Using the best-performing Parrotron model, we conducted
listening studies on a more challenging test set, which contains heavily accented
speech plus background noise. As shown in Table~\ref{tbl:listening_test}, we
verify that under these conditions Parrotron still preserves the linguistic
content, since its WER is comparable to that of real speech.
The naturalness MOS score decreases slightly with Parrotron when compared to that of real speech.
Recall that the objective in this work is to perform many-to-one speech normalization, not to improve ASR.
Training an ASR engine on the output of Parrotron is likely to improve WER results.
However, we leave evaluation of the impact of such normalization on ASR to future work.

\begin{table}[t]
\caption{
  \label{tbl:listening_study2}
  Subjective evaluation of Parrotron output quality.}
  \vskip-0.2cm
  \centering
  \begin{footnotesize}
  \begin{tabular}{l@{\hspace{0em}}r}
    \toprule
    Survey question & Avg. score / agreement \\
    \midrule
    How similar is the Parrotron voice to the \\
      \hspace{1em}
      TTS voice on the 5 point Likert scale? & 4.6 \\
    \addlinespace
    Does the output speech \\
    \hspace{1em}use a standard American English accent? & 94.4\%\\
    \hspace{1em}contain {\em any} background noise? & 0.0\%\\
    \hspace{1em}contain {\em any} disfluencies? & 0.0\%\\
    \hspace{1em}use consistent articulation, standard\\
    \hspace{2em}intonation and prosody? & 83.3\%\\
    \bottomrule
  \end{tabular}
  \end{footnotesize}
  \vskip-0.2cm
\end{table}

Finally, we conduct another listening test to evaluate whether the model consistently generates normalized speech with the same TTS voice.
We present a random sample of 20 utterances produced by Parrotron to 8 native English subjects and ask questions shown in Table~\ref{tbl:listening_study2}
for each utterance.
The results in the table verify that the model consistently normalizes speech. %

\subsubsection{Error analysis}

We analyze the types of phoneme errors Parrotron makes after normalization.
We first obtain the true phonemes by force aligning each manual
transcript with the corresponding real speech signal. Using this alignment, we
compute two confusion matrices on the test set: (A) %
one computed by aligning the true phonemes with
the hypothesized phonemes from the original speech, i.e.\ the Parrotron input;
(B) %
another computed by aligning the true phonemes to the
hypothesized phonemes from the normalized speech. We subtract A from B and rank the
phoneme confusions
to identify confusions which occur more frequently in Parrotron output than in real speech.
Since we have 40 phonemes (+ epsilon), we have 1681 phoneme
confusion pairs. %
In the top 5\% of confusions, we
observe that ~26\% of them are plosives (/k/, /t/, /d/, /g/, and /b/)
which are mostly dropped. The average rank of plosive confusions is 244/1681,
suggesting that the model does not accurately model these short phonemes. We also
observe another 12\% correspond to %
vowel exchanges. This is not surprising
since the model attempts to normalize multiple accents to that of the
target TTS speaker.

Errors in plosive and other short phonemes are not surprising since the model uses an L2 reconstruction loss. %
Under this loss, a frame containing a vowel contributes the same amount as a frame containing /t/.
Since there are significantly more vowel frames than plosives in the training data, this biases training to focus more on accurately generating phonemes of longer duration.

We observe that feeding Arabic and Spanish utterances into the US-English Parrotron model
often results in output which echoes the original speech content with an American accent, in the target voice.
Such %
behavior is qualitatively different from what one would obtain by
simply running an ASR followed by a TTS for example.
A careful listening study is needed to further validate these results.

\subsection{Normalization of hearing-impaired %
  speech}
\newcommand{\kadapt}{{\sc kadpt}}
\newcommand{\kdev}{{\sc kdev}}
\newcommand{\ktest}{{\sc ktest}}

Addressing a more challenging accessibility application, we investigate whether the normalization model can be used to
to convert atypical speech from a deaf speaker into fluent speech.
This could be used to improve the vocal communication of people with such conditions or other speech disorders,
or as a front-end to voice-enabled systems.

We focus on one case study of a profoundly deaf subject who
was born in Russia to normal-hearing parents, and
learned English as a teenager. %
The subject used Russian phonetic representation of English words and learned
to speak them using Russian letters (e.g., cat $\rightarrow$  k a T).
Using a
live (human in the loop) transcription service and ASR systems for multiple
years helped improve their articulation. See \cite{princtonstudy} for more details.

We experiment with adapting the best model from Section~\ref{sec:norm} using a dataset of 15.4~hours of speech, corresponding to read movie quotes.
We use 90\% of the data for adaptation ({\kadapt}), and hold out the remainder: %
 5\% (about 45 minutes) for dev and 5\% for test ({\ktest}).
This data was
challenging; we learned that some prompts %
were difficult to
pronounce by unimpaired but non-native
English speakers. The WER using Google's ASR
system on the TTS-synthesized %
{\em reference} transcripts is 14.8\%.
See the companion website for examples.

\subsubsection{Experiments}

\begin{table}[t]
\caption{
    \label{tbl:listening_test-dimitri}
    Performance on speech from a deaf speaker.} %
  \vskip-0.2cm
  \centering
  \begin{tabular}{lcr}
    \toprule
    Model & MOS & WER \\
    \midrule
    Real speech  & 2.08 $\pm$ 0.22 & 89.2 \\
    \midrule

    Parrotron (male) & 2.58 $\pm$ 0.20 & 109.3 \\
    Parrotron (male) finetuned %
                     & \bf{3.52 $\pm$ 0.14} & \bf{32.7} \\
    \bottomrule
  \end{tabular}
  \vskip-0.3cm
\end{table}

Our first experiment is to test the performance of Google's state-of-the-art ASR system on {\ktest}.
As shown in Table~\ref{tbl:listening_test-dimitri}, we find that the ASR system
performs very poorly on this speech, obtaining 89.2\% WER on the test set.
The MOS score on {\ktest} is 2.08, rated by subjects unfamiliar with the subject's
speech.

We then test whether our best out-of-the-box Parrotron trained for the normalization
task, shown in Section~\ref{sec:norm}, can successfully normalize this type of speech.
The only difference here is that Parrotron is trained on a male TTS speech, obtained form
our production WaveNet-based TTS. Testing on {\ktest}, we find that the output of this model was
rated as natural as the original speech, but our ASR engine performs even more poorly
on the converted speech than the original speech. In other words, Parrotron normalization
system trained on standard speech fails completely to normalize this type of speech.
We have also manually inspected the output of this Parrotron and found that the
model produces speech-like sounds but nonsense words.

Now, we test whether utilizing {\kadapt} would have any impact on Parrotron performance.
We first take the fully converged male Parrotron normalization model and
conduct multiple finetuning experiments using {\kadapt}. With a constant learning rate of 0.1, we
(1) adapt all parameters on the fully converged model;
(2) adapt all parameters except freezing the spectrogram decoder parameters;
(3) freeze both spectrogram decoder and phoneme decoder parameters while
finetuning only the encoder.

We find that all finetuning strategies %
lead to intelligible and significantly more natural speech.
The best finetuning strategy was adapting {\em all} parameters, which
increased the
MOS naturalness score by over 1.4 points compared to the original speech,
and dramatically reduced the WER from 89.2\% to 32.7\%.
Finetuning strategy (2) obtains 34.1\% WER and
adapting only encoder parameters (strategy (3)), obtains 38.6\% WER.

Note that one advantage of directly converting speech to speech
over cascading a finetuned ASR engine with TTS is as follows.
Synthesizing the output of an ASR engine may generate speech
far from intended, due to unavoidable ASR errors. A speech-to-speech
model, however, is likely to produce sounds closer to the original
speech. We have seen significant evidence to support this hypothesis,
but leave it to future work to quantify.

\subsection{Speech separation}

Finally, to illustrate that the Parrotron architecture can be used in a
variety of speech applications, we evaluate it on a speech separation
task of {\em reconstructing} the signal from the loudest speaker within a mixture
of overlapping speech.
We focus on instantaneous mixtures of up to 8 different speakers.

It is important to stress that our intent in this section is not to propose a state of the art
separation system, but rather to demonstrate
that the proposed architecture may apply to different speech applications.
More importantly, in contrast to previous applications which made use of
synthetic training targets, we evaluate whether Parrotron is able to generate speech from an
open set of speakers, generalizing beyond the training set.
Furthermore, unlike state-of-the-art speech separation
techniques \cite{hershey2016deep,wilson2018exploring}, Parrotron generates the signal
from scratch as opposed to using a masking-based filtering approach and is able to rely on
an implicit phoneme language model.

We use the same voice-search data described in
Section~\ref{sec:norm} to artificially construct instantaneous mixtures of speech signals.
For each target utterance %
in the training data, we randomly select a
set of 1 to 7 utterances to mix together as the background noise.
The number of
background utterances is also randomly selected. Before mixing, we normalize all utterances to have similar gains.

We mix target utterances with the background noise by simply averaging the two signals with a
randomly sampled weight  $w \in [0.1, 0.5)$ for the background
and $1-w$ for the target utterance. This results in an
average SNR across all artificially constructed utterances of 12.15 dB, with a standard
deviation of 4.7.
188K utterances from this corpus are held out for testing.
While we do not explicitly incorporate reverberation or non-speech noise,
the underlying utterances come from a variety of recording environments with their
own background noise. %

To evaluate whether Parrotron can perform this separation task, we
train a model to the best performing architecture as in
Section~\ref{sec:norm}. %
We feed as inputs our mixed utterances and train the model to generate
corresponding original {\em clean} utterances.

We evaluate the impact of this separation model using Google's ASR system.
We compare WERs on three sets of 188k held-out utterances:
(1) the original {\em clean} speech before
adding background speech; (2) the {\em noisy set} after mixing background speech;
(3) the cleaned output generated by running Parrotron
on the noisy set.
As shown in Table~\ref{table:separate}, we observe significant
WER reduction after running Parrotron on the noisy set, demonstrating that the model
can preserve speech from the target speaker and separate them
from other speakers.
Parrotron significantly reduces insertions, which correspond to words spoken by background speakers,
but suffers from increased deletions, which is likely due to early end-of-utterance prediction.

\begin{table}[t]
\caption{
    \label{table:separate}
    Parrotron speech separation performance.}
  \vskip-0.2cm
  \centering
  \begin{tabular}{lrrrr}
    \toprule
    Data & WER & del & ins & sub \\
    \midrule
    Original (Clean) & 8.8 & 1.6 & 1.5 & 5.8  \\
    Noisy & 33.2 & 3.6 & 19.1 & 10.5 \\
    Denoised using Parrotron & \bf{17.3} & 6.7 & 2.2 & 8.4 \\
    \bottomrule
  \end{tabular}

  \vskip-0.3cm
\end{table}

\section{Conclusion}
\label{sec:con}

We described Parrotron, an end-to-end speech-to-speech model that converts an input
spectrogram directly to another spectrogram, without intermediate symbolic
representation. We find that the model can be trained to normalize speech from
different speakers into speech of a single target speaker's voice while
preserving the linguistic content and projecting away non-linguistic content.
We then showed that this model can successfully be adapted to improve WER and
naturalness of speech from a deaf speaker. We finally demonstrate that the same
model can be trained to successfully identify, separate and
reconstruct the loudest speaker in a mixture of overlapping speech,
improving ASR performance.
The Parrotron system has other potential applications, e.g.\
improving intelligibility by converting heavily accented or otherwise atypical speech
into standard speech. In the future,  we plan to test it on other speech disorders,
and adopt techniques from \cite{jia2018transfer,haque2018conv} to  preserve the speaker identity.

\section{Acknowledgments}
We thank Françoise Beaufays, Michael Brenner, Diamantino Caseiro, Zhifeng Chen, Mohamed Elfeky, Patrick Nguyen, Bhuvana Ramabhadran, Andrew Rosenberg, Jason Pelecanos, Johan Schalkwyk, Yonghui Wu, and Zelin Wu for useful feedback.

\bibliographystyle{IEEEtran}
\bibliography{refs}

\end{document}